\documentclass[cameraready]{Interspeech}
\usepackage{amsmath}
\usepackage{multirow}
\usepackage{graphicx}
\usepackage{bm}
\usepackage{algorithm}
\usepackage{algorithmic}
 \usepackage{amsmath}
 \usepackage{booktabs}
 \usepackage{graphicx}
 \usepackage{hyperref}
 \usepackage{wasysym}
 \usepackage{multirow}
 \usepackage{booktabs}
 \usepackage{pifont}
 \usepackage{setspace}
\newcommand{\myparagraph}[1]{\vspace{6pt}\noindent\textbf{#1.}\quad}

\title{DGS-MLDG: Domain Gradient Surgery Guided Meta-Learning for Domain Generalization in Speech Deepfake Detection}

\author[affiliation={1}]{Siqing}{QIN}
\author[affiliation={1}, correspondingauthor]{Kong Aik}{LEE}
\author[affiliation={1}]{Youzhi}{TU}
\author[affiliation={2}]{Eng Siong}{CHNG}
\author[affiliation={1}]{Man-Wai}{MAK}

\address{
    $^1$ Dept. of Electrical and Electronic Engineering, The Hong Kong Polytechnic University \\
    $^2$ College of Computing and Data Science, Nanyang Technological University 
}

\email{siqing.qin@connect.polyu.hk}


\begin{document}

\maketitle
%
\begin{abstract}
Speech deepfake detection faces significant challenges due to domain shifts. Domain generalization (DG), particularly meta-learning for domain generalization (MLDG), offers a promising solution by simulating and mitigating domain shifts. However, MLDG is often hindered by conflicting gradients between its meta-train and meta-test objectives, leading to suboptimal performance. To address this problem, we propose domain gradient surgery (DGS), a meta-learning method that resolves conflicts through an asymmetric projection strategy. DGS removes the destructive component from the meta-test gradient, ensuring a conflict-free optimization trajectory versus the meta-train gradient. Furthermore, we introduce layer-wise DGS (LW-DGS), an efficient variant of DGS that dynamically identifies and intervenes only conflict-prone layers. Extensive experiments on challenging benchmarks demonstrate that DGS-MLDG and LW-DGS-MLDG achieve an average relative EER reduction of 5.29\% and 4.04\%, respectively.
\end{abstract}
%
%
\section{Introduction}
\label{sec:intro}
Speech deepfakes generated by advanced algorithms pose a severe threat to automatic speaker verification \cite{jin2025denoising, jin2026uncertainty} and public information security \cite{asvspoof2019, yamagishi2021asvspoof, asvspoof5, cfad2024, tran2025leveraging}. While deep learning detectors perform remarkably well in controlled environments, their reliability deteriorates significantly in real-world scenarios due to \textit{domain shifts} \cite{li2025survey}. This phenomenon arises from a mismatch between the training data and unseen test conditions, such as unseen attack types \cite{kwok2025bona, crs_ds3}. Unlike unsupervised domain adaptation (UDA), which requires access to target-domain data, domain generalization (DG) aims to learn domain-invariant representations from the source domains only, making it more practical in security applications \cite{huang2025shift, xie2023dg, ren2025dg}.

Standard DG approaches often rely on empirical risk minimization (ERM) by pooling all source domains \cite{vapnik1991erm, ge2025post}, but this strategy lacks explicit exploration in domain structures. To address this limitation, the \textit{meta-learning domain generalization} (MLDG) framework \cite{li2018mldg} has emerged as a promising alternative. By treating each attack type as a separate domain, MLDG simulates domain shifts during training through meta-train and meta-test splits. The aim is to mitigate such domain shifts by reducing the performance gap between the splits. Extensions such as multi-task meta-learning (MTML) have been proposed to stabilize the meta-optimization trajectory and help escape local minima \cite{metatrim2024}. In \cite{2025metageneralizable}, MLDG was combined with parameter-efficient tuning methods such as low-rank adaptation (LoRA) \cite{hu2022lora}. These approaches typically outperform ERM in cross-domain evaluations.

\begin{figure}
    \centering
    \includegraphics[width=0.8\linewidth]{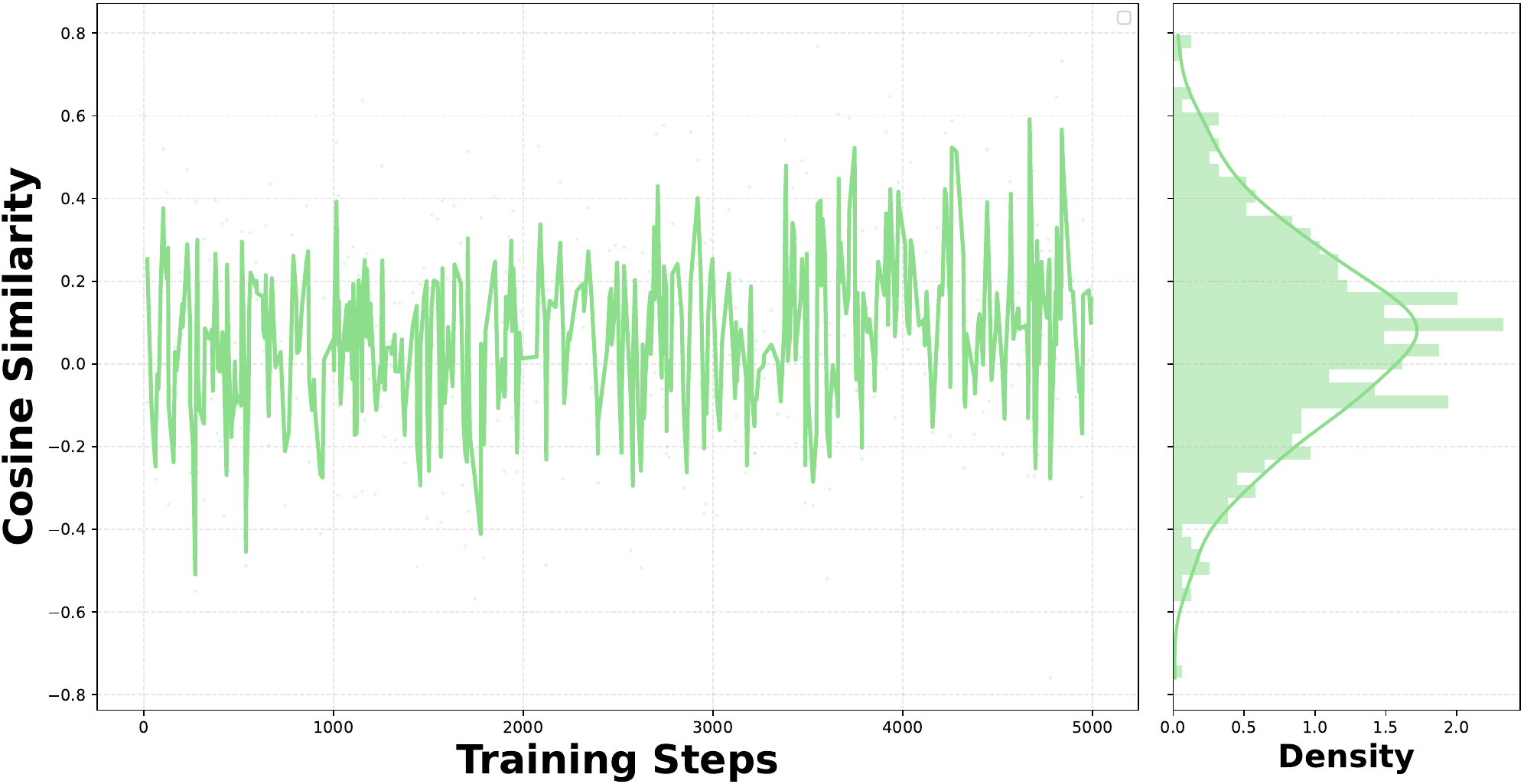}
    \caption{Negative cosine similarity between the meta-train gradient and meta-test gradient of MLDG during training.}
    \label{fig:mldgcos}
\end{figure}
However, applying MLDG to large-scale speech backbones introduces a critical optimization bottleneck known as \textit{gradient conflict}, which is a serious issue in multi-task learning (MTL) and DG \cite{yu2020gradient, mansilla2021domain, truong2025addressing}. Our analysis reveals that the gradients, derived from the meta-train and meta-test optimization of MLDG, frequently point in opposite directions, as shown in Figure \ref{fig:mldgcos}. The right panel shows the probability density (distribution) of cosine similarity values across all training steps, and higher density indicates values that occur more frequently. Simple aggregation of these gradients, also called naive gradient aggregation (NGA), can cancel out conflicting gradient accumulation, leading to destructive model optimization. On the other hand, gradient-based methods, such as PCGrad \cite{yu2020gradient}, GradVac\cite{wang2020gradient}, and CAGrad \cite{liu2021conflict}, are not directly applicable to the bi-level optimization of meta-learning, where preserving specific directional components is crucial.

\begin{figure*}[t]
    \centering
    \includegraphics[width=1.0\linewidth]{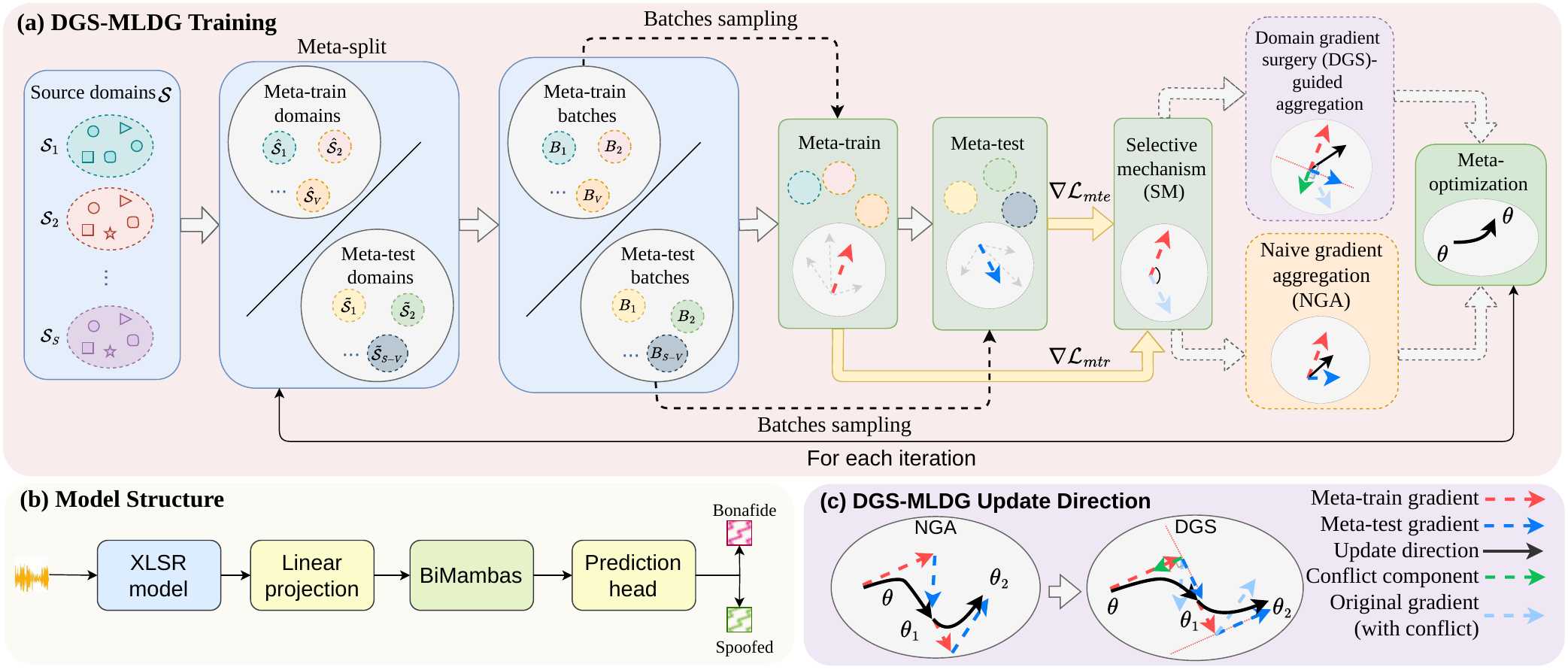}
    \caption{Overview of the proposed DGS-MLDG framework. (a) The training pipeline simulates domain shift via meta-splits and resolving conflicts via the selective mechanism (SM). (b) The backbone architecture used for deepfake detection. (c) Geometric comparison showing how DGS corrects the update direction by projecting conflicting gradients, avoiding the oscillation of Naive Gradient Aggregation (NGA).}
    \label{fig:meta}
\end{figure*}
To address the gradient conflict issue of MLDG, we propose \textbf{domain gradient surgery-guided MLDG (DGS-MLDG)}, a principled framework designed to resolve gradient conflicts in speech deepfake detection, as shown in Figures \ref{fig:meta}(a) and \ref{fig:meta}(c). Unlike NGA, DGS projects the conflicting meta-test gradient onto the normal plane of the meta-train gradient. This operation removes the destructive parameter update between the meta-train and meta-test optimization. 
Furthermore, we introduce \textbf{layer-wise DGS (LW-DGS-MLDG)} to learn domain-invariant and task-specific components based on layer-level gradient statistics. It is motivated by the observation that domain-specific information and artifact traces are associated with specific parameters, such as the affine parameters in normalization layer \cite{li2016bn, wu2024bn}, and with early layers of self-supervised learning (SSL) models \cite{el2025ssl}. 
LW-DGS applies surgery selectively to this conflict-prone subset determined by \textit{real-time cosine similarity monitoring}.

The main contributions of this paper are summarized as follows: (1) We provide an investigation of gradient dynamics in MLDG for speech deepfake detection, identifying gradient conflict between meta-train and meta-test objectives as a factor limiting generalization performance. (2) We propose the DGS-MLDG framework, which utilizes a novel projection mechanism to mitigate gradient conflicts. (3) We further develop a layer-wise strategy (LW-DGS) to address layer-level conflicts. Extensive experiments on multiple out-of-domain datasets demonstrate that our method achieves improvements against a strong ERM baseline. Evaluation covers state-of-the-art synthesis attacks, codec/transmission artifacts, and in-the-wild recordings.

\section{Proposed Methods}
\label{sec:meth}
This section introduces domain gradient surgery (DGS) and its efficient variant, layer-wise DGS (LW-DGS), to mediate the gradient conflict in MLDG.

\subsection{Preliminaries: Framework}
\label{subsec:prelim_mldg}
As illustrated in Figure \ref{fig:meta}(a), MLDG involves a meta-train (source adaptation) phase and a meta-test (generalization) phase. The source domains $\mathcal{S}=\{\mathcal{S}_i\}_{i=1}^{S}$ are partitioned into $V$ meta-train domains $\hat{\mathcal{S}}$ and $S-V$ meta-test domains $\tilde{\mathcal{S}}$ in each iteration. The meta-train loss $\mathcal{L}_{mtr}$ guides the inner-loop adaptation, while the meta-test loss $\mathcal{L}_{mte}$ evaluates the generalization on adapted parameters $\theta' = \theta - \alpha \nabla_\theta \mathcal{L}_{mtr}(\theta)$. The overall objective minimizes the weighted sum:
\begin{align}
    \mathop{\arg\min}_{\theta}\mathcal{L}_{mtr}(\theta) + \beta\mathcal{L}_{mte}(\theta - \alpha \nabla_\theta \mathcal{L}_{mtr}(\theta)),
\end{align}
where $\beta$ weights the meta-test phase. 
Standard MLDG uses naive gradient aggregation (NGA) to combine gradients linearly. NGA makes the model perform well on source domains while maintaining the capacity for fast adaptation to unseen domains. However, this often leads to conflicting gradients between $\mathcal{L}_{mtr}$ and $\mathcal{L}_{mte}$, hindering convergence.

\subsection{Domain Gradient Surgery (DGS)}
\label{subsec:dgs}
The gradients from the two phases, meta-train gradient $\bm{F} = \nabla\mathcal{L}_{mtr}$ and meta-test gradient $\bm{G} = \nabla\mathcal{L}_{mte}$, often conflict with each other, leading to unstable optimization as depicted in Figure \ref{fig:meta}(c). DGS uses an innovative selective mechanism to address the gradient conflict issue.

\textbf{Selective Mechanism (SM):} DGS monitors the inner product between $\bm{F}$ and $\bm{G}$. If $\langle \bm{G}, \bm{F} \rangle \geq 0$, gradients are combined directly. If $\langle \bm{G}, \bm{F} \rangle < 0$, signifying conflict, the DGS projection mechanism is enabled.

\textbf{Meta-Optimization with DGS:} When conflict is detected ($\langle \bm{G}, \bm{F} \rangle < 0$), DGS employs an asymmetric projection strategy. It treats $\bm{F}$ as the anchor for domain adaptation and removes only the detrimental component of $\bm{G}$ that deviates from $\bm{F}$:
\begin{align}
\label{equ:proj}
\bm{G}_{\small \text{proj}} = 
\begin{cases}
    \bm{G} - \left(\frac{\langle \bm{G}, \bm{F} \rangle}{||\bm{F}||^2 + \epsilon} \right) \bm{F}, \quad &\text{if } \langle \bm{G}, \bm{F} \rangle < 0,\\
    \bm{G}, &\text{otherwise}.
\end{cases}
\end{align}
Here, $\epsilon$ is a small constant (e.g., $10^{-8}$) for numerical stability. DGS ensures $\langle \bm{G}_{\small \text{proj}}, \bm{F} \rangle \geq 0$, guaranteeing a conflict-free gradient descent. The model parameters $\bm{\theta}$ are then updated:
\begin{equation}
\bm{\theta} \leftarrow \bm{\theta} - \gamma \left( \bm{F} + \beta \bm{G}_{\small \text{proj}} \right).
\end{equation}
This asymmetric strategy is crucial for stable meta-optimization, as it explicitly preserves the essential domain-specific knowledge provided by $\bm{F}$ for inner-loop adaptation, allowing $\bm{G}_{\small \text{proj}}$ to guide generalization effectively. In contrast, NGA may lead to suboptimal oscillating updates (Figure \ref{fig:meta}(c)). This mechanism distinguishes DGS from multi-task gradient-based surgery methods such as PCGrad \cite{yu2020gradient}. The complete procedure is in Algorithm \ref{alg:modified_mldg}.

\subsection{Layer-Wise Domain Gradient Surgery (LW-DGS)}
\label{subsec:LW_dgs}

Global DGS can be computationally intensive for large backbones (e.g., 300M-parameter SSL backbones). Furthermore, as observed in \cite{li2016bn,wu2024bn,el2025ssl}, some parameters, such as normalization layers and early SSL layers, are more sensitive to domain-specific statistics and artifact traces. This motivates LW-DGS, an efficient approach that performs surgery only where conflicts genuinely occur at the layer level.

\textbf{Layer-Wise Selective Mechanism (LW-SM):} LW-DGS performs \textit{independent conflict detection} at the layer level. Let $\bm{\theta}_k$ denote a specific trainable layer (e.g., a weight matrix) within the model's total parameter set $\bm{\theta}$. For each parameter $\bm{\theta}_k$, we consider its corresponding meta-train gradient tensor $\bm{F}_k = \nabla_{\bm{\theta}_k}\mathcal{L}_{mtr}(\bm{\theta})$ and meta-test gradient tensor $\bm{G}_k = \nabla_{\bm{\theta}_k}\mathcal{L}_{mte}(\bm{\theta}')$. 
We then compute their individual cosine similarity:
\begin{equation}
\text{cos}_{\bm{\theta}_k} = \frac{\langle \bm{F}_k, \bm{G}_k \rangle}{||\bm{F}_k|| \cdot ||\bm{G}_k||}.
\end{equation}

\textbf{Dynamic Surgery:} 
The selective modification is then applied \textit{dynamically and independently} to each parameter based on its instantaneous conflict status. 
When gradient tensors $\bm{F}_k$ and $\bm{G}_k$ are found in conflict (i.e., $\langle \bm{F}_k, \bm{G}_k \rangle < 0$), the component of $\bm{G}_k$ that opposes $\bm{F}_k$ is adjusted.
\begin{equation}
\label{eq:LW_proj}
\bm{G}_k^{\small \text{proj}} = 
\begin{cases}
\bm{G}_k - \frac{\langle \bm{G}_k, \bm{F}_k \rangle}{||\bm{F}_k||^2 + \epsilon} \bm{F}_k, & \text{if } \langle \bm{F}_k, \bm{G}_k \rangle < 0 \\
\bm{G}_k, & \text{otherwise}.
\end{cases}
\end{equation}
Here, $\epsilon$ is a constant (e.g., $10^{-8}$) for numerical stability. This operation effectively suppresses the conflict of $\bm{G}_k$ with $\bm{F}_k$ at the layer level, ensuring that the modified gradient $\bm{G}_k^{\small \text{proj}}$ contributes constructively to the update direction.

This fine-grained, layer-level monitoring provides a significant advantage by precisely identifying and intervening only on components most susceptible to domain shifts. 
Conversely, domain-insensitive parameters are inherently optimized for the core binary classification task in each minibatch. Therefore, LW-DGS naturally bypasses these insensitive parameters and focuses on the parameters that are likely to face gradient conflict only, saving computational resources. 

\begin{algorithm}[t]
\caption{DGS-MLDG for Speech Deepfake Detection}
\label{alg:modified_mldg}
\setstretch{0.7}
\begin{algorithmic}[1]
\STATE \textbf{Input}: Domains $\mathcal{S}$
\STATE \textbf{Init}: Model parameters $\theta$. Hyperparameters $\alpha$, $\beta$, $\gamma$.
\FOR{each iteration}
    \STATE \textbf{Split}: $\mathcal{\hat{S}}$ and $\mathcal{\tilde{S}} \leftarrow \mathcal{S}$
    \STATE Initialize accumulators: $\bm{F} \leftarrow \bm{0}$, $\bm{G} \leftarrow \bm{0}$
    \STATE \textbf{Meta-train:} 
    \FOR{minibatch $\mathcal{B}_{\small \mathcal{\hat{S}}_{i}}$ from $\mathcal{\hat{S}}_i$} 
    \STATE\quad Compute gradient \(\bm F_{i}=\nabla_{\bm \theta}\mathcal{L}_{mtr}(\bm{\theta}, \mathcal{B}_{\mathcal{\hat{S}}_{i}})\)
    \STATE\quad Accumulate gradient $\bm{F} \leftarrow \bm{F} + \frac{1}{V}\bm{F}_i$
    \ENDFOR
    \STATE \textbf{Inner-update:} \(\bm{\theta}' \leftarrow \bm{\theta} - \alpha\bm{F}\)

    \STATE \textbf{Meta-test:}
    \FOR{minibatch \(\mathcal{B}_{\mathcal{\tilde{S}}_{j}}\)  from $\mathcal{\tilde{S}}$}
    
    \STATE\quad Compute gradient \(\bm{G}_j = \nabla_{\bm{\theta}} \mathcal{L}_{mte}(\bm{\theta}';\mathcal{B}_{\mathcal{\tilde{S}}_{j}})\)
    \STATE\quad Accumulate gradient $\bm{G} \leftarrow \bm{G} + \frac{1}{S-V}\bm{G}_j$
    \ENDFOR
    \STATE \textbf{DGS:}
    \IF{$\langle\bm{G}, \bm{F}\rangle < 0$}
    \STATE $\bm{G} = \bm{G} - \left( \frac{\langle\bm{G}, \bm{F}\rangle}{||\bm{F}||^2 +\epsilon} \right) \bm{F}$
    \ENDIF
    \STATE \textbf{Meta-optimization:} 
    \(\bm{\theta} \leftarrow \bm{\theta} - \gamma \times \frac{1}{2}  (\bm{F} + \beta \bm{G})\)
\ENDFOR
\end{algorithmic}
\end{algorithm}

\section{Experiments and results}
\label{sec:exp}
\subsection{Experimental Conditions}
\myparagraph{Training settings}
We selected three datasets for training: ASVspoof 2019 LA \cite{asvspoof2019} (\textbf{ASV2019}), ASVspoof 5 \cite{asvspoof5}, and CFAD \cite{cfad2024}. The training processes employed both the original training sets and development sets under varied conditions.
For naive joint-training minimizing empirical risk (i.e., ERM), we concatenate the three training sets to train the model collectively.
For meta-learning, we utilized the three datasets to define the source domains, resulting in a total of 22 domains: 6 attack types from the ASVspoof 2019 LA training set, 8 attack types from the ASVspoof 5 (\textbf{ASV5}) training set, and 8 attack types from the CFAD training set. 

\myparagraph{Implementation details} The model structure follows the design proposed in \cite{xiao2025xlsr}, as shown in Figure \ref{fig:meta}(b). 
To enhance data variability, RawBoost \cite{rawboost} was employed to introduce stationary, signal-independent additive noise. During training, audio segments were cropped or concatenated to approximately 4 seconds (64,600 samples). The Adam optimizer, with a learning rate of \(10^{-6}\) and a weight decay of \(10^{-4}\), was utilized to optimize the weighted cross-entropy loss with a batch size of 12. An inner learning rate of $10^{-2}$ and a $\beta$ of 0.5, determined by the validation set, yields optimal performance. 

\begin{table*}[t!]
\centering
\caption{Comparison of EER (\%) among main methods. The last column shows the cross-dataset average relative improvement w.r.t. the ERM baseline ($\uparrow$). Bold and underline denote the lowest and the second-lowest EERs for each dataset.}
\resizebox{1.0\linewidth}{!}{
\begin{tabular}{l|cccccccc}
\toprule
\textbf{Methods} & \textbf{ASV21}$\downarrow$ & \textbf{ASV5}$\downarrow$ & \textbf{CFAD}$\downarrow$ & \textbf{ADD2023 R1}$\downarrow$ & \textbf{ADD2023 R2}$\downarrow$ & \textbf{In-the-wild}$\downarrow$ & \textbf{CodecFake}$\downarrow$ & \textbf{Avg. Rel. Imp. ($\%$) $\uparrow$}\\
\hline
ERM & \textbf{0.92} & \textbf{8.57} & 24.33 & 19.16 & 23.23 & 6.71 & 7.66 & - \\
MLDG & 1.05 & 9.69 & 24.16 & 18.14 & 23.03 & 6.47 & \underline{7.24} & 2.53 \\ 

\hline
DGS-MLDG (ours) & \underline{1.00} & 10.24 & \underline{23.86} & \textbf{17.76} & 23.19 & \textbf{5.23} & \textbf{6.76} & \textbf{5.29} \\
LW-DGS-MLDG (ours) & 1.02 & \underline{9.55} & \textbf{23.83} & \underline{17.88} & \textbf{22.68} & \underline{6.15} & 7.27 & \underline{4.04} \\
\bottomrule
\end{tabular}
}
\label{tab:main_eer_comparison}
\end{table*}
\begin{table*}[t!]
\centering
\caption{Ablation study: EER (\%) under all evaluation settings.}
\resizebox{0.80\linewidth}{!}{
\begin{tabular}{lcccccccc}
\toprule
Experiments & \textbf{ASV21}$\downarrow$ & \textbf{ASV5}$\downarrow$ & \textbf{CFAD}$\downarrow$ & \textbf{ADD2023 R1}$\downarrow$ & \textbf{ADD2023 R2}$\downarrow$ & \textbf{In-the-wild}$\downarrow$ & \textbf{CodecFake}$\downarrow$\\ 
\midrule
Ablation \ding{172} & 0.97 & 9.82 & 24.27 & 20.43 & 23.68 & 6.15 & 7.27  \\ 
Ablation \ding{173} & 1.16 & 9.57 & 23.93 & 18.03 & 22.75 & 6.58 & 7.27  \\ 
\bottomrule
\end{tabular}
}
\label{tab:ablation_eer_comparison}
\end{table*}
\myparagraph{Evaluation settings}
For in-dataset evaluation, we used the ASVspoof 2021 DF (\textbf{ASV21}) test set \cite{yamagishi2021asvspoof}, and the ASVspoof 5 test set (\textbf{ASV5}). 
For cross-dataset evaluation, we selected four additional datasets: the \textbf{ADD 2023} \cite{yi2023add} test sets (comprising Test Set R1 and Test Set R2), the \textbf{In-the-wild} \cite{muller2022itw}, the \textbf{CodecFake} \cite{codecfake} test sets. The CodecFake evaluation comprises 7 subsets, and we calculated the mean EER across them. 
We treated the CFAD unseen test set (\textbf{CFAD}) as a cross-dataset evaluation set because the fake attack types and real speech sources do not overlap in it. 
\vspace{-0.1cm}
\subsection{Results and Analysis}
\label{subsec:results}
\begin{figure}
    \centering
    \includegraphics[width=0.8\linewidth]{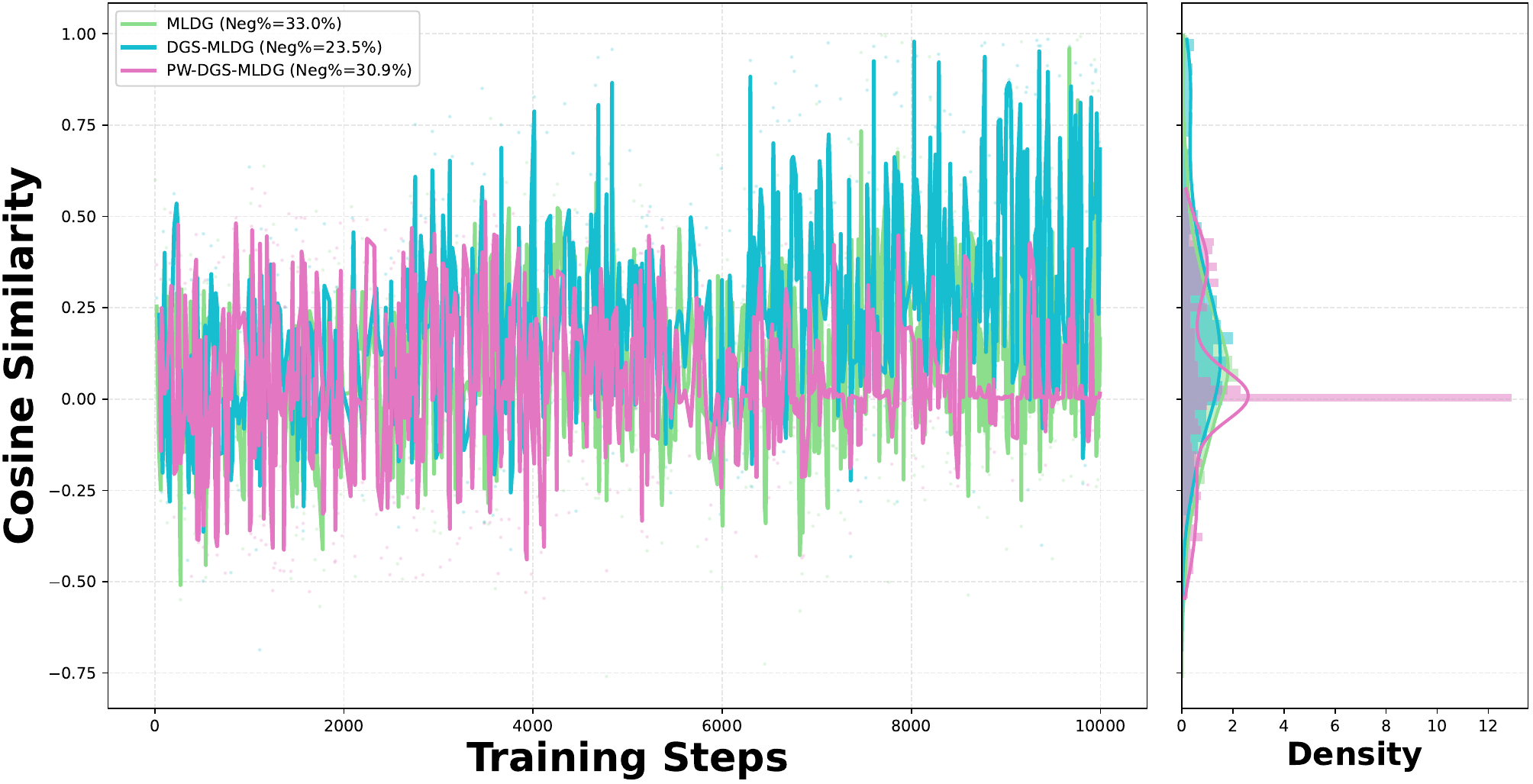}
    \caption{Cosine similarity comparison across MLDG, DGS-MLDG, and LW-DGS-MLDG. Conflicts decrease via proposed DGS and LW-DGS.}
    \label{fig:coscom}
\end{figure}

\myparagraph{Gradient Alignment Analysis} We first analyze the evolution of gradient dynamics during training. As illustrated in Figure \ref{fig:coscom}, the cosine similarity between meta-train and meta-test gradients in our proposed DGS framework exhibits a consistent upward trend, converging to high positive values. We observe that DGS-MLDG is concentrated around slightly higher cosine similarity across all training steps. In contrast, the standard MLDG baseline shows fluctuating low similarity. This observation shows that DGS effectively alleviates gradient conflicts, aligning source-domain adaptation with cross-domain generalization objectives.

\myparagraph{Cross-Dataset Generalization Performance} 
Table \ref{tab:main_eer_comparison} presents Equal Error Rate (EER) results on cross-dataset benchmarks. The ERM baseline consistently yields high error rates, particularly on distinct domains like TIMIT-TTS and In-the-Wild, confirming its inability to learn domain-invariant representations. While the standard MLDG provides marginal improvements over ERM, it still suffers from optimization instability due to conflicting gradients.
\begin{table}[t]
\centering
\caption{Comparison of EER (\%) results across different gradient alignment methods.}
\label{tab:comparison_sota}
\resizebox{0.80\linewidth}{!}{\begin{tabular}{lcc}
\toprule
\textbf{System} 
 & \textbf{In-the-wild} & \textbf{CodecFake}\\ \midrule
XLSR-Mamba & 6.71 & 7.66  \\
\hspace{3mm} w/ MLDG training & 6.47 & 7.24 \\
\hspace{6mm} + DGS (ours) & \textbf{5.23} & \textbf{6.76}\\
\hspace{6mm} + LW-DGS (ours) & 6.15 & 7.27 \\
\hspace{6mm} + PCGrad & 6.17 & 7.69 \\
\hspace{6mm} + GradVac & 6.34 & 7.40 \\
\hspace{6mm} + CAGrad & 7.40&7.36 \\
\bottomrule
\end{tabular}}
\label{tab:grad}
\end{table}
Our DGS-MLDG framework significantly outperforms all baselines, achieving the lowest EERs across nearly all cross-dataset evaluation sets. Specifically, it attains a relative performance improvement of \textbf{5.29\%} over the strongest baseline. The gains are particularly pronounced on challenging datasets such as ADD2023 R1, CodecFake, and In-the-Wild, demonstrating better generalization against unseen codecs and real-world spoofing scenarios. Furthermore, the efficient LW-DGS-MLDG achieves a comparable relative improvement of \textbf{4.04\%}, validating that resolving conflicts in a small subset of sensitive parameters is sufficient to obtain substantial generalization gains. These results confirm that removing conflicting components allows a stable meta-optimization under rigorous domain shifts.

\subsection{Comparison with Gradient Alignment Methods}
\vspace{-0.1cm}
In Table \ref{tab:grad}, our proposed methods are compared with general multi-task gradient surgery techniques, including PCGrad \cite{yu2020gradient}, GradVac \cite{wang2020gradient}, and CAGrad \cite{liu2021conflict}. A critical observation is that these methods, which employ bidirectional gradient surgery (treating tasks symmetrically), fail to consistently outperform the standard MLDG baseline. For instance, PCGrad degrades on CodecFake. This suggests that treating the meta-train and meta-test objectives as equal peers disrupts the domain knowledge learned during the meta-train phase.

In contrast, our proposed DGS and LW-DGS maintain consistent gradient directions between two phases during parameter update. This asymmetric strategy follows the bi-level optimization of meta-learning, preserving the primary adaptation signal while filtering out destructive interference from the generalization objective. Consequently, DGS achieves the lowest EERs on In-the-wild and CodecFake, confirming that the asymmetric strategy is essential for robust domain generalization.
\vspace{-0.1cm}
\subsection{Ablation Studies}
\vspace{-0.2cm}
\label{subsec:ablation}
\myparagraph{Impact of Projection Direction} 
We investigate another direction for resolving gradient conflicts by comparing DGS with its variants. Ablation \ding{172} uses the \textit{Reverse Projection}, where the conflicting meta-train gradient is projected onto the normal plane of the meta-test gradient. 
As shown in Table \ref{tab:ablation_eer_comparison}, a variant significantly underperforms the proposed DGS. The degradation in \textit{Reverse Projection} confirms that the meta-train gradient encodes domain-specific knowledge essential for inner-loop adaptation; modifying it according to the meta-test objective compromises the model's ability to learn basic discriminative features. 

\myparagraph{Effectiveness of Layer-Wise Monitoring} 
We further verify the necessity of monitoring the entire model in LW-DGS. 
We conduct an experiment (Ablation \ding{173}) that restricts gradient surgery to the pre-trained SSL backbone, leaving the downstream BiMamba backend untouched. 
Results in Table \ref{tab:ablation_eer_comparison} show that it results in higher EERs. 
This validates the advantage of our LW-DGS mechanism, which dynamically identifies and addresses conflicts across \textit{all} model parameters without imposing artificial architectural constraints.
\vspace{-0.1cm}
\section{Conclusion}
\vspace{-0.1cm}
This paper addressed the critical challenge of domain generalization in speech deepfake detection. We identified that while MLDG offers a promising paradigm, its effectiveness is often limited by conflicting gradients between meta-train and meta-test objectives. To overcome this, we proposed domain gradient surgery (DGS), a novel framework that systematically resolves these conflicts through an asymmetric gradient projection strategy. Its efficient variant, layer-wise DGS (LW-DGS), further enhances scalability by dynamically targeting only conflict-prone layers. Extensive experiments on challenging benchmarks demonstrated significant reductions in EER.

\section{Acknowledgment}
This work was supported in part by the Innovation and Technology Fund of the Hong Kong SAR (Project No. MHP/048/24), and the National Key R\&D Program of China (2024YFE0217200).

\section{Use of Generative AI Disclosure}
\label{GenAI}
Generative AI tools were used only for language polishing and formatting assistance. All scientific content, experiments, and analyses were produced and verified by the authors.

\bibliographystyle{IEEEtran}
\bibliography{mybib}

\end{document}